\documentclass[aps,12pt,superscriptaddress,amsfonts,amssymb,amsmath]{revtex4-2}
\pdfoutput=1

\usepackage{graphicx}
\usepackage{xcolor}
\usepackage{amsmath}
\usepackage{bm}
\usepackage{amssymb}
\usepackage[a4paper,margin=2.6cm]{geometry}
\usepackage{hyperref}
\usepackage{float}
\usepackage{pgfplots}
\usepgfplotslibrary{groupplots}
\pgfplotsset{compat=1.18}

\begin{document}

\title{
Statistics of non-conserved observables in Lindblad master equations
}

\author{G.\ Modanese \footnote{Email address: giovanni.modanese@unibz.it}}
\affiliation{Free University of Bozen-Bolzano \\ Faculty of Engineering \\ I-39100 Bolzano, Italy}
\date{\today}

\linespread{0.9}

\begin{abstract}
We study the dynamics of observables that are conserved under the Hamiltonian evolution of a closed quantum system, but cease to be conserved when the system is coupled to a Markovian environment and described by a Lindblad master equation. Starting from the adjoint Lindblad equation, we derive elementary expressions for the time derivatives of the expectation value and second moment of an observable $O$, with particular emphasis on the case $[H,O]=0$ but $\mathcal L^\dagger(O)\neq 0$. These formulae provide a direct assessment of how collapse operators break Hamiltonian conservation laws and generate fluctuations of formerly conserved quantities. The discussion is illustrated by analytic examples: one-qubit amplitude damping, a two-qubit excitation-number model, a momentum-diffusion model in which the mean is conserved while the variance grows, and the Jaynes--Cummings model. The latter also shows the complementary case of a reservoir coupled through a conserved quantity, where dephasing can occur without changing the statistics of that quantity. We finally comment on the relation between Lindblad source terms and idealized wave-function reduction models in which local conservation may hold only statistically.
\end{abstract}

\maketitle

\section{Introduction}
\label{sec:introduction}

Conservation laws play a central role in the analysis of closed quantum
systems. If a system evolves unitarily under a Hamiltonian $H$, an
observable $O$ is conserved whenever it commutes with the Hamiltonian,
\begin{equation}
    [H,O]=0 .
\end{equation}
Equivalently, its expectation value is constant for all initial states,
\begin{equation}
    \frac{d}{dt}\langle O\rangle =0 .
\end{equation}
This elementary statement is no longer sufficient for open quantum
systems. In the presence of an environment, the reduced density matrix of
the system may obey a Lindblad master equation, and the dissipative part
of the generator can break a conservation law that is present in the
Hamiltonian sector~\cite{Gorini1976,Lindblad1976,BreuerPetruccione2002,Manzano2020}.

The purpose of this work is to discuss this loss of conservation in a
simple and operational way. We consider observables $O$ which are
conserved by the Hamiltonian evolution, but not by the full Lindblad
dynamics. Rather than focusing only on the expectation value
$\langle O\rangle$, we also consider the second moment
$\langle O^2\rangle$ and the variance
\begin{equation}
    \mathrm{Var}(O)
    =
    \langle O^2\rangle-\langle O\rangle^2 .
\end{equation}
This gives a minimal description of the time-dependent statistics of an
observable which would be conserved in the corresponding closed system.
In this sense, $\mathcal D^\dagger(O)$ will be interpreted below as the
open-system source term associated with the observable $O$.

The relation between symmetries and conserved quantities in Lindblad
master equations has been analyzed in detail by Albert and Jiang
\cite{AlbertJiang2014}. One of the important lessons of that work is that
the usual unitary correspondence between continuous symmetries and
conserved quantities is modified in dissipative dynamics. In the present
paper we address a narrower question: given a simple Hamiltonian
conservation law, how do the expectation value, second moment, and
variance of the corresponding observable evolve once Lindblad collapse
operators are added?

It is important to stress that the aim of this paper is not to provide a
new classification of invariant operators or of stationary subspaces of a
Lindblad semigroup. Such structural questions are treated in greater
generality in the literature on quantum dynamical semigroups and
dissipative symmetries~\cite{Frigerio1978,BaumgartnerNarnhofer2008,BucaProsen2012}.
Our more modest goal is to isolate a simple moment-based diagnostic for
the transient statistics of a Hamiltonian conserved observable once it is
subjected to dissipative dynamics. In this sense the paper should be read
as complementary to the rigorous theory of Lindblad symmetries: it
focuses on the drift, second moment, and variance generated by the
dissipative part of the adjoint generator, in analytically transparent
models.

Our discussion is intentionally elementary. We first derive the formal
identities for the time derivatives of $\langle O\rangle$,
$\langle O^2\rangle$, and $\mathrm{Var}(O)$. We then illustrate them with
two simple examples: a one-qubit amplitude damping model and a two-qubit
model in which the total excitation number is conserved by the Hamiltonian
but not by local decay processes.

We also add a momentum-diffusion example, which realizes explicitly the
case in which the mean value of an observable is conserved while its
variance grows. Finally, we discuss the Jaynes--Cummings model, where the closed Hamiltonian
has a conserved total excitation number. This example is useful because
it shows both possibilities: a dephasing reservoir coupled through the
conserved excitation number preserves its statistics, whereas photon loss
or atomic decay makes the same quantity non-conserved.

A further motivation for studying not only the first moment but also the
second moment is that, in stochastic or reduction-like descriptions, a
conservation law may hold only after statistical averaging. In the
language used below, the non-Hamiltonian part of the evolution can act as
an effective source or sink for the observable under consideration. For a
local density this would correspond to the analogue of a source term in a
continuity equation; for a general observable it simply means that
$\mathcal D^\dagger(O)$ is nonzero. In such situations the average
source contribution to $\dot{\langle O\rangle}$ may vanish, whereas
fluctuations, second moments, or correlations can still carry information
about the underlying stochastic processes. This point is relevant, for
example, in idealized wave-function-reduction models in which local
charge conservation can be violated event by event while being restored
after statistical averaging in suitable situations~\cite{MinottiModanese2026Collapse}.
The Lindblad framework offers a simple setting in which the analogous
distinction can be analyzed at the level of observables and their
moments.

The paper is organized as follows. In Sec.~\ref{sec:lindblad-adjoint} we
recall the Lindblad equation and its adjoint form, emphasizing the
condition for an observable to be conserved by the full open dynamics. In
Sec.~\ref{sec:second-moment} we derive the evolution equations for the
second moment and the variance. In Sec.~\ref{sec:initial-values} we
summarize the role of the initial statistics of the observable. Sections
\ref{sec:one-qubit} and~\ref{sec:two-qubit} present two elementary spin
examples, including the explicit solution of the two-qubit moment
equations. Section~\ref{sec:variance-only} gives a simple
variance-only example based on momentum diffusion. Section~\ref{sec:jaynes-cummings} discusses the
Jaynes--Cummings excitation number and contrasts dephasing through a
conserved quantity with dissipative channels that destroy it. In
Sec.~\ref{sec:collapse-reduction} we comment on the relation with local
non-conservation in wave-function-reduction models. The main conclusions
are collected in Sec.~\ref{sec:conclusions}.

\section{Lindblad dynamics and adjoint evolution}
\label{sec:lindblad-adjoint}

We consider a finite-dimensional quantum system whose density matrix
$\rho(t)$ evolves according to a time-independent Lindblad master
equation,
\begin{equation}
    \dot{\rho}
    =
    \mathcal{L}(\rho)
    =
    -i[H,\rho]
    +
    \sum_k \gamma_k
    \left(
        L_k \rho L_k^\dagger
        -
        \frac{1}{2}
        \left\{
            L_k^\dagger L_k,\rho
        \right\}
    \right),
    \label{eq:lindblad}
\end{equation}
where we set $\hbar=1$. Here $H=H^\dagger$ is the system Hamiltonian,
$L_k$ are the Lindblad or collapse operators, and $\gamma_k\geq 0$ are
the corresponding rates. This is the standard time-homogeneous GKSL form
of a Markovian quantum master equation~\cite{Gorini1976,Lindblad1976,Carmichael1993,GardinerZoller2004,WisemanMilburn2010,RivasHuelga2012}.

For an observable $O=O^\dagger$, the expectation value is
\begin{equation}
    \langle O\rangle(t)
    =
    \mathrm{Tr}\bigl(\rho(t) O\bigr).
\end{equation}
Taking the time derivative and using Eq.~\eqref{eq:lindblad}, one finds
\begin{equation}
    \frac{d}{dt}\langle O\rangle
    =
    \mathrm{Tr}\bigl(\dot{\rho} O\bigr)
    =
    \mathrm{Tr}\bigl(\mathcal{L}(\rho) O\bigr).
\end{equation}
Using the cyclic property of the trace, this can be written as
\begin{equation}
    \frac{d}{dt}\langle O\rangle
    =
    \mathrm{Tr}\bigl(\rho\,\mathcal{L}^\dagger(O)\bigr)
    =
    \left\langle
        \mathcal{L}^\dagger(O)
    \right\rangle ,
    \label{eq:expectation_adjoint}
\end{equation}
where $\mathcal{L}^\dagger$ is the adjoint Lindblad generator,
\begin{equation}
    \mathcal{L}^\dagger(O)
    =
    i[H,O]
    +
    \sum_k \gamma_k
    \left(
        L_k^\dagger O L_k
        -
        \frac{1}{2}
        \left\{
            L_k^\dagger L_k,O
        \right\}
    \right).
    \label{eq:adjoint_lindblad}
\end{equation}
Thus the condition for $O$ to be conserved for all states is not simply
$[H,O]=0$, but
\begin{equation}
    \mathcal{L}^\dagger(O)=0 .
    \label{eq:lindblad_conservation}
\end{equation}

It is useful to separate the Hamiltonian and dissipative contributions,
\begin{equation}
    \mathcal{L}^\dagger(O)
    =
    i[H,O]
    +
    \mathcal{D}^\dagger(O),
\end{equation}
with
\begin{equation}
    \mathcal{D}^\dagger(O)
    =
    \sum_k \gamma_k
    \left(
        L_k^\dagger O L_k
        -
        \frac{1}{2}
        \left\{
            L_k^\dagger L_k,O
        \right\}
    \right).
    \label{eq:adjoint_dissipator}
\end{equation}
If $[H,O]=0$, then $O$ is conserved in the closed system. In the open
system, however,
\begin{equation}
    \frac{d}{dt}\langle O\rangle
    =
    \left\langle
        \mathcal{D}^\dagger(O)
    \right\rangle .
    \label{eq:dissipative_drift}
\end{equation}
Therefore the collapse operators determine whether and how the
Hamiltonian conservation law is broken.

A sufficient, though not necessary, condition for preservation of the
observable under the dissipative dynamics is
\begin{equation}
    [O,L_k]=[O,L_k^\dagger]=0
    \qquad
    \text{for all } k .
\end{equation}
In that case also
\begin{equation}
    [O,L_k^\dagger L_k]=0 ,
\end{equation}
and the dissipative contribution in Eq.~\eqref{eq:adjoint_dissipator}
vanishes. In general, however, conservation is controlled by the full
condition \eqref{eq:lindblad_conservation}.

\section{Second moment and variance}
\label{sec:second-moment}

The same formalism applies to any function of the observable. In
particular, for the second moment one has
\begin{equation}
    \langle O^2\rangle(t)
    =
    \mathrm{Tr}\bigl(\rho(t) O^2\bigr),
\end{equation}
and therefore
\begin{equation}
    \frac{d}{dt}\langle O^2\rangle
    =
    \left\langle
        \mathcal{L}^\dagger(O^2)
    \right\rangle .
    \label{eq:second_moment_general}
\end{equation}
Explicitly,
\begin{equation}
    \frac{d}{dt}\langle O^2\rangle
    =
    i\left\langle [H,O^2]\right\rangle
    +
    \sum_k \gamma_k
    \left\langle
        L_k^\dagger O^2 L_k
        -
        \frac{1}{2}
        \left\{
            L_k^\dagger L_k,O^2
        \right\}
    \right\rangle .
    \label{eq:second_moment_explicit}
\end{equation}
If $[H,O]=0$, then also $[H,O^2]=0$, and the evolution of the second
moment is entirely due to the dissipative part.

The variance is
\begin{equation}
    \mathrm{Var}(O)
    =
    \langle O^2\rangle-\langle O\rangle^2 .
\end{equation}
Taking the time derivative gives
\begin{equation}
    \frac{d}{dt}\mathrm{Var}(O)
    =
    \frac{d}{dt}\langle O^2\rangle
    -
    2\langle O\rangle
    \frac{d}{dt}\langle O\rangle .
\end{equation}
Using Eqs.~\eqref{eq:expectation_adjoint} and
\eqref{eq:second_moment_general}, this becomes
\begin{equation}
    \boxed{
    \frac{d}{dt}\mathrm{Var}(O)
    =
    \left\langle
        \mathcal{L}^\dagger(O^2)
    \right\rangle
    -
    2\langle O\rangle
    \left\langle
        \mathcal{L}^\dagger(O)
    \right\rangle
    } .
    \label{eq:variance_derivative}
\end{equation}
If, in particular, $[H,O]=0$, then also $[H,O^2]=0$, and Eq.~\eqref{eq:variance_derivative} reduces to
\begin{equation}
    \frac{d}{dt}\mathrm{Var}(O)
    =
    \left\langle
        \mathcal D^\dagger(O^2)
    \right\rangle
    -
    2\langle O\rangle
    \left\langle
        \mathcal D^\dagger(O)
    \right\rangle .
\end{equation}

It is useful to give this quantity a name. We shall refer to
\begin{equation}
    \Pi_O(t)
    =
    \left\langle
        \mathcal L^\dagger(O^2)
    \right\rangle
    -
    2\langle O\rangle
    \left\langle
        \mathcal L^\dagger(O)
    \right\rangle
    \label{eq:variance_production}
\end{equation}
as the variance-production rate of the observable $O$. For a Hamiltonian
conserved observable, $[H,O]=0$, this rate is generated entirely by the
dissipative part of the adjoint generator,
\begin{equation}
    \Pi_O(t)
    =
    \left\langle
        \mathcal D^\dagger(O^2)
    \right\rangle
    -
    2\langle O\rangle
    \left\langle
        \mathcal D^\dagger(O)
    \right\rangle .
    \label{eq:variance_production_dissipative}
\end{equation}
The content of Eq.~\eqref{eq:variance_derivative} is therefore not the
chain rule itself, but the separation between dissipative drift of the
mean and dissipative production or suppression of fluctuations.

This identity is the basic tool used in the following. It shows
that the loss of conservation of $O$ can manifest itself both as a drift
of the mean value and as a change in the width of the distribution.

There are several possible cases. If
\begin{equation}
    \mathcal{L}^\dagger(O)=0,
    \qquad
    \mathcal{L}^\dagger(O^2)=0,
\end{equation}
then both the expectation value and the variance of $O$ are conserved.
If
\begin{equation}
    \mathcal{L}^\dagger(O)\neq 0,
\end{equation}
then the mean value is generally not conserved. Finally, the case
\begin{equation}
    \mathcal{L}^\dagger(O)=0,
    \qquad
    \mathcal{L}^\dagger(O^2)\neq 0
\end{equation}
would correspond to conservation of the mean value but not of the
fluctuations. This distinction is useful because conservation of an
expectation value does not necessarily imply conservation of the full
statistics of the observable.

\section{Initial values and interpretation}
\label{sec:initial-values}

The formal derivatives derived above have to be interpreted together
with the initial statistics of $O$. Given an initial density matrix
$\rho_0$, one has
\begin{equation}
    \langle O\rangle_0
    =
    \mathrm{Tr}(\rho_0 O),
    \qquad
    \langle O^2\rangle_0
    =
    \mathrm{Tr}(\rho_0 O^2),
\end{equation}
and
\begin{equation}
    \mathrm{Var}(O)_0
    =
    \langle O^2\rangle_0-\langle O\rangle_0^2 .
\end{equation}

If the initial state is an eigenstate of $O$,
\begin{equation}
    O|\psi_0\rangle=o_0|\psi_0\rangle,
    \qquad
    \rho_0=|\psi_0\rangle\langle\psi_0|,
\end{equation}
then
\begin{equation}
    \langle O\rangle_0=o_0,
    \qquad
    \langle O^2\rangle_0=o_0^2,
    \qquad
    \mathrm{Var}(O)_0=0 .
\end{equation}
In this case any subsequent nonzero variance is generated dynamically by
the open-system evolution.

On the other hand, if the initial state is a coherent superposition of
different eigenstates of $O$,
\begin{equation}
    |\psi_0\rangle
    =
    \sum_n c_n |o_n\rangle,
    \qquad
    O|o_n\rangle=o_n|o_n\rangle,
\end{equation}
then
\begin{equation}
    \langle O\rangle_0
    =
    \sum_n |c_n|^2 o_n,
    \qquad
    \langle O^2\rangle_0
    =
    \sum_n |c_n|^2 o_n^2,
\end{equation}
and the variance is already nonzero unless the superposition contains
only one eigenvalue of $O$. Similarly, for a mixed initial state,
\begin{equation}
    \rho_0=\sum_\alpha p_\alpha
    |\psi_\alpha\rangle\langle\psi_\alpha|,
\end{equation}
the initial variance contains both quantum and statistical contributions.

Thus the dynamics of a formerly conserved quantity should be described
not only by whether $\langle O\rangle$ changes, but also by whether the
state starts from a sharp or broad distribution of $O$.

\section{A variance-only example: momentum diffusion}
\label{sec:variance-only}

The distinction between conservation of the mean and conservation of the
full statistics can be illustrated by a simple diffusion model. Consider a
particle with Hamiltonian
\begin{equation}
    H=\frac{p^2}{2m},
    \label{eq:diffusion_hamiltonian}
\end{equation}
and take as observable
\begin{equation}
    O=p.
\end{equation}
The Hamiltonian part conserves $p$, since $[H,p]=0$. Let the dissipative
part be generated by a Lindblad operator proportional to position,
\begin{equation}
    L=\sqrt{\eta}\,x,
    \label{eq:position_lindblad}
\end{equation}
where $\eta>0$. The corresponding dissipator is
\begin{equation}
    \mathcal D(\rho)
    =
    \eta
    \left(
        x\rho x-\frac12\{x^2,\rho\}
    \right),
    \label{eq:position_decoherence_dissipator}
\end{equation}
and its adjoint action can be written in the double-commutator form
\begin{equation}
    \mathcal D^\dagger(A)
    =
    -\frac{\eta}{2}[x,[x,A]].
    \label{eq:double_commutator}
\end{equation}
Such double-commutator dissipators are standard in simple models of
position decoherence and momentum diffusion~\cite{JoosZeh1985,BreuerPetruccione2002}.

For the first moment one has
\begin{equation}
    [x,[x,p]]=0,
\end{equation}
and hence
\begin{equation}
    \mathcal L^\dagger(p)=0.
    \label{eq:momentum_mean_conserved}
\end{equation}
Thus the mean momentum is conserved. However, using $[x,p]=i$ with
$\hbar=1$,
\begin{equation}
    [x,p^2]=2ip,
    \qquad
    [x,[x,p^2]]=-2,
\end{equation}
so that
\begin{equation}
    \mathcal L^\dagger(p^2)=\eta.
    \label{eq:momentum_second_moment_growth}
\end{equation}
Consequently,
\begin{equation}
    \frac{d}{dt}\langle p\rangle=0,
    \qquad
    \frac{d}{dt}\mathrm{Var}(p)=\eta.
    \label{eq:momentum_variance_growth}
\end{equation}
This example realizes explicitly the case
$\mathcal L^\dagger(O)=0$ but $\mathcal L^\dagger(O^2)\neq 0$. The
dissipative dynamics produces no drift of the mean momentum, but it
produces a linear growth of momentum fluctuations.

\subsection{Relation with fluctuation--dissipation balance}

The momentum-diffusion example also clarifies the relation between the
present moment-based description and fluctuation--dissipation ideas. A
generic Lindblad generator need not satisfy a fluctuation--dissipation
theorem: this requires additional physical input, such as a thermal
environment, a detailed-balance condition, or a prescribed thermal
stationary state. Nevertheless, the quantities considered here are
precisely those which enter such a balance.

In the pure diffusion model considered above, the mean momentum is
conserved, while the variance grows according to
\begin{equation}
    \frac{d}{dt}\mathrm{Var}(p)=\eta .
\end{equation}
This represents the fluctuation-producing part of the dynamics without
a compensating dissipative relaxation. In a thermal Brownian model one
would also have a friction term, schematically
\begin{equation}
    \frac{d}{dt}\langle p\rangle=-\Gamma\langle p\rangle,
\end{equation}
and
\begin{equation}
    \frac{d}{dt}\langle p^2\rangle
    =
    -2\Gamma\langle p^2\rangle+2D_p .
\end{equation}
For states with \(\langle p\rangle=0\), this gives
\begin{equation}
    \frac{d}{dt}\mathrm{Var}(p)
    =
    -2\Gamma\,\mathrm{Var}(p)+2D_p .
\end{equation}
At equilibrium the left-hand side vanishes, so that
\begin{equation}
    \mathrm{Var}_{\rm eq}(p)=\frac{D_p}{\Gamma}.
\end{equation}
If the stationary state is thermal in the high-temperature limit,
\(\mathrm{Var}_{\rm eq}(p)=mk_BT\), and therefore
\begin{equation}
    D_p=\Gamma m k_BT .
\end{equation}
This is the usual fluctuation--dissipation balance in moment form:
the same environmental coupling that produces damping also fixes the
strength of the fluctuations required to reach thermal equilibrium.
From this perspective, the variance-production term
\begin{equation}
    \Pi_O(t)
    =
    \left\langle\mathcal L^\dagger(O^2)\right\rangle
    -
    2\langle O\rangle
    \left\langle\mathcal L^\dagger(O)\right\rangle
\end{equation}
is the natural object through which fluctuation--dissipation relations
appear at the level of a chosen observable.

Thus the present formalism does not by itself imply a
fluctuation--dissipation theorem for arbitrary collapse operators.
However, when the Lindblad generator has a thermal origin, the equations
for the first and second moments provide the direct route by which the
dissipative drift and the fluctuation-producing terms are related.

\section{One-qubit case: amplitude damping}
\label{sec:one-qubit}

As a first example, consider a two-level system with Hamiltonian
\begin{equation}
    H
    =
    \frac{\omega}{2}\sigma_z .
\end{equation}
We take
\begin{equation}
    O=\sigma_z .
\end{equation}
Then
\begin{equation}
    [H,\sigma_z]=0,
\end{equation}
so $\sigma_z$ is conserved in the closed system.

We now add a single amplitude-damping collapse operator, a standard elementary dissipative channel in open two-level systems~\cite{Manzano2020},
\begin{equation}
    L=\sigma_-,
    \qquad
    \gamma>0.
\end{equation}
We use the convention
\begin{equation}
    \sigma_z
    =
    |1\rangle\langle 1|
    -
    |0\rangle\langle 0|,
    \qquad
    \sigma_-
    =
    |0\rangle\langle 1|,
    \qquad
    \sigma_+
    =
    |1\rangle\langle 0|.
\end{equation}
Thus $|1\rangle$ is the excited state and $|0\rangle$ the ground state.

The adjoint dissipator acting on $\sigma_z$ is
\begin{equation}
    \mathcal{D}^\dagger(\sigma_z)
    =
    \gamma
    \left(
        \sigma_+ \sigma_z \sigma_-
        -
        \frac{1}{2}
        \left\{
            \sigma_+\sigma_-,
            \sigma_z
        \right\}
    \right).
\end{equation}
Since
\begin{equation}
    \sigma_+\sigma_-
    =
    |1\rangle\langle 1|,
\end{equation}
one obtains
\begin{equation}
    \sigma_+ \sigma_z \sigma_-
    =
    - |1\rangle\langle 1|,
\end{equation}
and
\begin{equation}
    \left\{
        \sigma_+\sigma_-,
        \sigma_z
    \right\}
    =
    2 |1\rangle\langle 1| .
\end{equation}
Therefore
\begin{equation}
    \mathcal{D}^\dagger(\sigma_z)
    =
    -2\gamma |1\rangle\langle 1| .
\end{equation}
Using
\begin{equation}
    |1\rangle\langle 1|
    =
    \frac{1}{2}
    \left(
        \mathbb{I}+\sigma_z
    \right),
\end{equation}
we get
\begin{equation}
    \mathcal{D}^\dagger(\sigma_z)
    =
    -\gamma
    \left(
        \mathbb{I}+\sigma_z
    \right).
\end{equation}
Thus
\begin{equation}
    \frac{d}{dt}\langle\sigma_z\rangle
    =
    -\gamma
    \left(
        1+\langle\sigma_z\rangle
    \right).
    \label{eq:sigmaz_decay}
\end{equation}
The solution is
\begin{equation}
    \langle\sigma_z\rangle(t)
    =
    -1+
    \left(
        1+\langle\sigma_z\rangle_0
    \right)
    e^{-\gamma t}.
    \label{eq:sigmaz_solution}
\end{equation}
For an initially excited state,
\begin{equation}
    \langle\sigma_z\rangle_0=1,
\end{equation}
this gives
\begin{equation}
    \langle\sigma_z\rangle(t)
    =
    -1+2e^{-\gamma t}.
\end{equation}

The second moment is trivial in this case, because
\begin{equation}
    \sigma_z^2=\mathbb{I}.
\end{equation}
Therefore
\begin{equation}
    \langle\sigma_z^2\rangle=1
\end{equation}
for all normalized states and at all times. Consequently,
\begin{equation}
    \mathrm{Var}(\sigma_z)
    =
    1-\langle\sigma_z\rangle^2 .
\end{equation}
For the initially excited state,
\begin{equation}
    \mathrm{Var}(\sigma_z)(t)
    =
    1-
    \left(
        -1+2e^{-\gamma t}
    \right)^2 .
\end{equation}
The variance starts from zero, becomes nonzero during the decay process,
and returns to zero asymptotically as the system approaches the ground
state.

This example is analytically transparent, but it is also somewhat
special because $O^2=\mathbb{I}$. For this reason the second moment does
not provide independent information. A richer example is obtained by
considering a two-qubit system with a conserved excitation number.

\section{Two-qubit case: conserved excitation number}
\label{sec:two-qubit}

We now consider two qubits with an exchange Hamiltonian,
\begin{equation}
    H_{exch}
    =
    g
    \left(
        \sigma_+^{(1)}\sigma_-^{(2)}
        +
        \sigma_-^{(1)}\sigma_+^{(2)}
    \right).
    \label{eq:exchange_hamiltonian}
\end{equation}
Equivalently, up to conventional factors,
\begin{equation}
    H_{exch}
    =
    \frac{g}{2}
    \left(
        \sigma_x^{(1)}\sigma_x^{(2)}
        +
        \sigma_y^{(1)}\sigma_y^{(2)}
    \right).
\end{equation}
The total excitation number is
\begin{equation}
    N
    =
    n_1+n_2,
    \qquad
    n_j
    =
    \sigma_+^{(j)}\sigma_-^{(j)}
    =
    |1\rangle_j\langle 1|_j .
    \label{eq:number_operator}
\end{equation}
The operator $N$ has eigenvalues $0,1,2$, corresponding respectively to
the states
\begin{equation}
    |00\rangle,
    \qquad
    |10\rangle,\ |01\rangle,
    \qquad
    |11\rangle .
\end{equation}

The Hamiltonian \eqref{eq:exchange_hamiltonian} conserves the total
excitation number. Indeed, using
\begin{equation}
    [n_j,\sigma_+^{(j)}]=\sigma_+^{(j)},
    \qquad
    [n_j,\sigma_-^{(j)}]=-\sigma_-^{(j)},
\end{equation}
and the fact that operators acting on different qubits commute, one
finds
\begin{align}
    [N,\sigma_+^{(1)}\sigma_-^{(2)}]
    &=
    [n_1+n_2,\sigma_+^{(1)}\sigma_-^{(2)}]
    \nonumber \\
    &=
    [n_1,\sigma_+^{(1)}]\sigma_-^{(2)}
    +
    \sigma_+^{(1)}[n_2,\sigma_-^{(2)}]
    \nonumber \\
    &=
    \sigma_+^{(1)}\sigma_-^{(2)}
    -
    \sigma_+^{(1)}\sigma_-^{(2)}
    =
    0 .
\end{align}
Similarly,
\begin{equation}
    [N,\sigma_-^{(1)}\sigma_+^{(2)}]=0 .
\end{equation}
Therefore
\begin{equation}
    [H_{exch},N]=0 .
\end{equation}
In the closed system,
\begin{equation}
    \frac{d}{dt}\langle N\rangle=0,
    \qquad
    \frac{d}{dt}\langle N^2\rangle=0,
    \qquad
    \frac{d}{dt}\mathrm{Var}(N)=0 .
\end{equation}

We now add local (i.e., acting separately on the two qubits) amplitude-damping collapse operators,
\begin{equation}
    L_1=\sigma_-^{(1)},
    \qquad
    L_2=\sigma_-^{(2)},
\end{equation}
with rates $\gamma_1$ and $\gamma_2$. The dissipator acting on the density matrix is
\begin{equation}
    \mathcal{D}(\rho)
    =
    \sum_{j=1}^2
    \gamma_j
    \left(
        \sigma_-^{(j)}\rho\sigma_+^{(j)}
        -
        \frac{1}{2}
        \left\{
            n_j,\rho
        \right\}
    \right).
\end{equation}
The adjoint dissipator is
\begin{equation}
    \mathcal{D}^\dagger(O)
    =
    \sum_{j=1}^2
    \gamma_j
    \left(
        \sigma_+^{(j)} O \sigma_-^{(j)}
        -
        \frac{1}{2}
        \left\{
            n_j,O
        \right\}
    \right).
\end{equation}

We first compute the evolution of the mean excitation number. Since
\begin{equation}
    N=n_1+n_2,
\end{equation}
and the decay of qubit $j$ only changes $n_j$, one obtains
\begin{equation}
    \mathcal{D}^\dagger(N)
    =
    -\gamma_1 n_1-\gamma_2 n_2 .
    \label{eq:Ddagger_N_twoqubits}
\end{equation}
Therefore
\begin{equation}
    \boxed{
    \frac{d}{dt}\langle N\rangle
    =
    -\gamma_1 \langle n_1\rangle
    -
    \gamma_2 \langle n_2\rangle
    } .
    \label{eq:N_mean_twoqubits}
\end{equation}
For equal rates,
\begin{equation}
    \gamma_1=\gamma_2=\gamma,
\end{equation}
this reduces to
\begin{equation}
    \frac{d}{dt}\langle N\rangle
    =
    -\gamma \langle N\rangle .
\end{equation}
Hence
\begin{equation}
    \langle N\rangle(t)
    =
    \langle N\rangle_0 e^{-\gamma t}
\end{equation}
for equal local decay rates.

The second moment is more informative. Since $n_j^2=n_j$,
\begin{equation}
    N^2
    =
    (n_1+n_2)^2
    =
    n_1+n_2+2n_1n_2
    =
    N+2n_1n_2 .
    \label{eq:N2_twoqubits}
\end{equation}
The operator $n_1n_2$ projects onto the doubly excited state
$|11\rangle$. Equation~\eqref{eq:N2_twoqubits} shows that this sector contributes
specifically to the second moment through the term $2n_1n_2$.
A local decay maps $|11\rangle$ into one of the one-excitation
states, $|10\rangle$ or $|01\rangle$, so that $N$ changes from
$2$ to $1$ and $N^2$ changes from $4$ to $1$. This is the origin
of the $n_1n_2$ contribution in the evolution equation for
$\langle N^2\rangle$ below.

A direct application of the adjoint dissipator gives
\begin{equation}
    \mathcal{D}^\dagger(N^2)
    =
    -\gamma_1 n_1(1+2n_2)
    -
    \gamma_2 n_2(1+2n_1).
    \label{eq:Ddagger_N2_twoqubits}
\end{equation}
Equivalently,
\begin{equation}
    \mathcal{D}^\dagger(N^2)
    =
    -\gamma_1 n_1
    -
    \gamma_2 n_2
    -
    2(\gamma_1+\gamma_2)n_1n_2 .
\end{equation}
Thus
\begin{equation}
    \boxed{
    \frac{d}{dt}\langle N^2\rangle
    =
    -\gamma_1 \langle n_1\rangle
    -
    \gamma_2 \langle n_2\rangle
    -
    2(\gamma_1+\gamma_2)\langle n_1 n_2\rangle
    } .
    \label{eq:N2_mean_twoqubits}
\end{equation}
For equal decay rates,
\begin{equation}
    \gamma_1=\gamma_2=\gamma,
\end{equation}
this becomes
\begin{equation}
    \frac{d}{dt}\langle N^2\rangle
    =
    -\gamma\langle N\rangle
    -
    4\gamma\langle n_1n_2\rangle .
\end{equation}
Using Eq.~\eqref{eq:N2_twoqubits},
\begin{equation}
    \langle n_1n_2\rangle
    =
    \frac{1}{2}
    \left(
        \langle N^2\rangle-\langle N\rangle
    \right),
\end{equation}
we get
\begin{equation}
    \frac{d}{dt}\langle N^2\rangle
    =
    \gamma\langle N\rangle
    -
    2\gamma\langle N^2\rangle .
    \label{eq:N2_equal_rates}
\end{equation}
Together with
\begin{equation}
    \frac{d}{dt}\langle N\rangle
    =
    -\gamma\langle N\rangle ,
\end{equation}
this gives a closed system of equations for the first and second moments.

The variance obeys
\begin{equation}
    \frac{d}{dt}\mathrm{Var}(N)
    =
    \frac{d}{dt}\langle N^2\rangle
    -
    2\langle N\rangle
    \frac{d}{dt}\langle N\rangle .
\end{equation}
For equal rates, using the previous equations,
\begin{equation}
    \boxed{
    \frac{d}{dt}\mathrm{Var}(N)
    =
    \gamma\langle N\rangle
    -
    2\gamma\langle N^2\rangle
    +
    2\gamma\langle N\rangle^2
    } .
    \label{eq:variance_N_twoqubits}
\end{equation}

For equal rates the closed system of moment equations can be solved
explicitly. Let
\begin{equation}
    N_0=\langle N\rangle_0,
    \qquad
    M_0=\langle N^2\rangle_0 .
    \label{eq:initial_moments_twoqubit}
\end{equation}
Then
\begin{equation}
    \langle N\rangle(t)=N_0e^{-\gamma t},
    \label{eq:N_solution_twoqubit}
\end{equation}
and
\begin{equation}
    \langle N^2\rangle(t)
    =
    N_0e^{-\gamma t}
    +
    (M_0-N_0)e^{-2\gamma t}.
    \label{eq:N2_solution_twoqubit}
\end{equation}
Therefore
\begin{equation}
    \mathrm{Var}(N)(t)
    =
    N_0e^{-\gamma t}
    +
    (M_0-N_0)e^{-2\gamma t}
    -
    N_0^2e^{-2\gamma t}.
    \label{eq:variance_solution_twoqubit_general}
\end{equation}
For an initial eigenstate of $N$ with eigenvalue $n_0$, so that
$N_0=n_0$ and $M_0=n_0^2$, this reduces to
\begin{equation}
    \mathrm{Var}(N)(t)
    =
    n_0e^{-\gamma t}
    \left(
        1-e^{-\gamma t}
    \right).
    \label{eq:variance_solution_twoqubit_eigenstate}
\end{equation}
In particular, for the initially doubly excited state $|11\rangle$,
\begin{equation}
    \mathrm{Var}(N)(t)
    =
    2e^{-\gamma t}
    \left(
        1-e^{-\gamma t}
    \right),
    \label{eq:variance_solution_twoqubit_11}
\end{equation}
which reaches its maximum value $1/2$ at $t=(\ln 2)/\gamma$.

\begin{figure}[H]
    \centering
    \begin{tikzpicture}
        \begin{axis}[
            width=0.55\linewidth,
            height=0.34\linewidth,
            xlabel={$\gamma t$},
            ylabel={$\mathrm{Var}(N)$},
            xmin=0, xmax=5,
            ymin=0, ymax=0.55,
            samples=200,
            domain=0:5,
            grid=both
        ]
        \addplot[blue, thick] {2*exp(-x)*(1-exp(-x))};
        \end{axis}
    \end{tikzpicture}
    \caption{Dynamical generation of the excitation-number variance for
    the initial state $|11\rangle$ in the equal-rate two-qubit model,
    Eq.~\eqref{eq:variance_solution_twoqubit_11}.}
    \label{fig:variance_twoqubit_11}
\end{figure}
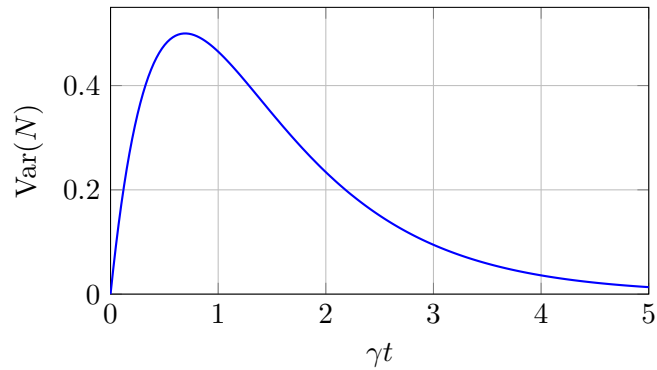

The equal-rate assumption is essential for the closure of the moment
equations in terms of $\langle N\rangle$ and $\langle N^2\rangle$ alone.
For unequal rates, Eq.~\eqref{eq:N_mean_twoqubits} depends separately
on $\langle n_1\rangle$ and $\langle n_2\rangle$. Their Hamiltonian
evolution is coupled to the exchange coherence. For example, defining
\begin{equation}
    Y
    =
    i\left\langle
        \sigma_-^{(1)}\sigma_+^{(2)}
        -
        \sigma_+^{(1)}\sigma_-^{(2)}
    \right\rangle,
    \label{eq:exchange_coherence_Y}
\end{equation}
one obtains schematically
\begin{equation}
    \frac{d}{dt}\langle n_1\rangle
    =
    gY-\gamma_1\langle n_1\rangle,
    \qquad
    \frac{d}{dt}\langle n_2\rangle
    =
    -gY-\gamma_2\langle n_2\rangle,
    \label{eq:unequal_rate_populations}
\end{equation}
together with an additional equation for $Y$. Thus, for
$\gamma_1\neq\gamma_2$, the population moments do not form a closed
two-equation system. The equal-rate case is therefore the minimal case in
which the moment dynamics of $N$ and $N^2$ closes analytically.

This model displays more structure than the one-qubit example. In the
closed system, the Hamiltonian exchanges an excitation between the two
qubits but preserves the total number $N$. The local collapse operators,
instead, connect different excitation sectors,
\begin{equation}
    |11\rangle
    \longrightarrow
    |10\rangle,\ |01\rangle,
    \qquad
    |10\rangle,\ |01\rangle
    \longrightarrow
    |00\rangle .
\end{equation}
Thus the same observable that is exactly conserved by the Hamiltonian
becomes a fluctuating, decaying quantity in the Lindblad dynamics.

For example, if the system is initially in the state $|11\rangle$, then
\begin{equation}
    \langle N\rangle_0=2,
    \qquad
    \langle N^2\rangle_0=4,
    \qquad
    \mathrm{Var}(N)_0=0 .
\end{equation}
The subsequent dynamics generates a distribution over the sectors
$N=2,1,0$, and the variance becomes nonzero at intermediate times.

If instead the initial state is in the one-excitation subspace, for
example
\begin{equation}
    |\psi_0\rangle
    =
    \frac{1}{\sqrt{2}}
    \left(
        |10\rangle+|01\rangle
    \right),
\end{equation}
then
\begin{equation}
    \langle N\rangle_0=1,
    \qquad
    \langle N^2\rangle_0=1,
    \qquad
    \mathrm{Var}(N)_0=0 .
\end{equation}
The exchange Hamiltonian preserves this subspace, but the local decay
operators transfer population to $|00\rangle$. Therefore the variance of
$N$ is again dynamically generated by the dissipative evolution.

These examples illustrate the central point: a quantity may be exactly
conserved by the Hamiltonian and nevertheless acquire nontrivial
statistics under Lindblad evolution. The first moment measures the
dissipative drift of the quantity, while the second moment and variance
measure the broadening or narrowing of its distribution.

\section{Jaynes--Cummings model: conserved and non-conserved excitation number}
\label{sec:jaynes-cummings}

A standard physical example of a Hamiltonian conserved quantity is provided
by the Jaynes--Cummings model~\cite{JaynesCummings1963}. In the rotating-wave approximation, the
Hamiltonian of a two-level atom coupled to a single quantized mode is
\begin{equation}
    H_{\rm JC}
    =
    \omega_0 a^\dagger a
    +
    \frac{\varepsilon}{2}\sigma_z
    +
    g\left(
        a\sigma_+ + a^\dagger\sigma_-
    \right),
    \label{eq:H_JC}
\end{equation}
where $a,a^\dagger$ are the annihilation and creation operators of the
field mode, while $\sigma_\pm$ and $\sigma_z$ act on the two-level atom.
The interaction term describes the absorption of a photon accompanied by
atomic excitation, and the emission of a photon accompanied by atomic
de-excitation.

The total excitation number is
\begin{equation}
    N
    =
    a^\dagger a+\sigma_+\sigma_- .
    \label{eq:N_JC}
\end{equation}
The Jaynes--Cummings Hamiltonian conserves $N$,
\begin{equation}
    [H_{\rm JC},N]=0.
\end{equation}
Indeed, the operator $a\sigma_+$ annihilates one photon and excites the
atom, while $a^\dagger\sigma_-$ creates one photon and de-excites the
atom. In both processes the total number of excitations is unchanged.

This conserved quantity is closely related to the solvability of the
closed Jaynes--Cummings model. The Hilbert space decomposes into invariant
subspaces of fixed $N$, each of which is two-dimensional, apart from the
ground sector. This allows the model to be solved analytically. In open
versions of the model, however, the fate of this conserved quantity
depends on the structure of the system--environment coupling.

A useful contrast is given by a pure-dephasing coupling through the
conserved quantity $N$ itself. In the exact open Jaynes--Cummings model
studied by Watanabe and Nakazato~\cite{WatanabeNakazato2022}, the reservoir interaction is mediated
by $N$. Although the general discussion above was introduced for a
time-homogeneous generator, the adjoint identity extends directly to
time-local generators. If
\begin{equation}
    \dot\rho(t)=\mathcal L_t(\rho(t)),
    \label{eq:time_local_generator}
\end{equation}
then for a time-independent observable $O$ one has
\begin{equation}
    \frac{d}{dt}\langle O\rangle
    =
    \left\langle \mathcal L_t^\dagger(O)\right\rangle .
    \label{eq:time_local_adjoint}
\end{equation}
Thus the same adjoint reasoning applies also to the exact time-local
master equation used in this example. The resulting reduced master equation has a dissipative term of
the form
\begin{equation}
    \dot\rho
    =
    -i[H_{\rm JC}+\dot f_t N^2,\rho]
    +
    2\dot h_t
    \left(
        N\rho N
        -
        \frac{1}{2}\{N^2,\rho\}
    \right),
    \label{eq:Watanabe_master}
\end{equation}
with time-dependent coefficients. In this case the Lindblad operator is
effectively proportional to $N$, and therefore
\begin{equation}
    \mathcal L^\dagger(N)=0,
    \qquad
    \mathcal L^\dagger(N^2)=0.
\end{equation}
Thus the reservoir can destroy coherences between different $N$ sectors,
but it does not change the statistics of $N$ itself.

The situation is different for the more usual dissipative channels of
cavity photon loss and atomic spontaneous emission. These may be modeled
phenomenologically by the collapse operators
\begin{equation}
    L_a=a,
    \qquad
    L_\sigma=\sigma_-,
\end{equation}
with rates $\kappa$ and $\gamma$, respectively. The dissipator is
\begin{equation}
    \mathcal D(\rho)
    =
    \kappa
    \left(
        a\rho a^\dagger
        -
        \frac{1}{2}\{a^\dagger a,\rho\}
    \right)
    +
    \gamma
    \left(
        \sigma_-\rho\sigma_+
        -
        \frac{1}{2}\{\sigma_+\sigma_-,\rho\}
    \right).
    \label{eq:JC_dissipator_loss}
\end{equation}
Let
\begin{equation}
    n=a^\dagger a,
    \qquad
    e=\sigma_+\sigma_-,
    \qquad
    N=n+e.
\end{equation}
The adjoint dissipator gives
\begin{equation}
    \mathcal D^\dagger(N)
    =
    -\kappa n-\gamma e.
    \label{eq:Ddagger_N_JC}
\end{equation}
Therefore
\begin{equation}
    \frac{d}{dt}\langle N\rangle
    =
    -\kappa\langle n\rangle
    -
    \gamma\langle e\rangle .
    \label{eq:N_JC_mean}
\end{equation}
Although $N$ is conserved by the Hamiltonian, it is not conserved by the
open dynamics when photons can leak out of the cavity or the atom can
decay into external modes.

The second moment is also simple to analyze. Since $e^2=e$ and $n$
commutes with $e$,
\begin{equation}
    N^2
    =
    n^2+e+2ne.
\end{equation}
A photon-loss event changes $N$ into $N-1$ with rate proportional to
$n$, while an atomic decay event changes $N$ into $N-1$ with rate
proportional to $e$. Hence
\begin{equation}
    \mathcal D^\dagger(N^2)
    =
    \kappa n\left[(N-1)^2-N^2\right]
    +
    \gamma e\left[(N-1)^2-N^2\right].
\end{equation}
Equivalently,
\begin{equation}
    \mathcal D^\dagger(N^2)
    =
    (1-2N)(\kappa n+\gamma e).
    \label{eq:Ddagger_N2_JC}
\end{equation}
Thus
\begin{equation}
    \frac{d}{dt}\langle N^2\rangle
    =
    \left\langle
        (1-2N)(\kappa n+\gamma e)
    \right\rangle .
    \label{eq:N2_JC_mean}
\end{equation}
The variance of the total excitation number obeys
\begin{equation}
    \frac{d}{dt}\mathrm{Var}(N)
    =
    \left\langle
        (1-2N)(\kappa n+\gamma e)
    \right\rangle
    +
    2\langle N\rangle
    \left(
        \kappa\langle n\rangle+\gamma\langle e\rangle
    \right).
    \label{eq:Var_N_JC}
\end{equation}

This example shows explicitly the distinction between Hamiltonian
conservation and conservation under the full Lindblad generator. The
Jaynes--Cummings Hamiltonian preserves the total excitation number, but
standard dissipative processes such as cavity loss and atomic decay do
not. Conversely, a reservoir coupled through $N$ itself produces
dephasing without changing the first and second moments of $N$. The same
closed-system conserved quantity can therefore be either preserved or
destroyed by the environment, depending on the structure of the collapse
operators.

\section{Relation with local conservation and wave-function reduction}
\label{sec:collapse-reduction}

The discussion above concerns observables which are conserved by the
Hamiltonian part of the dynamics but not necessarily by the full Lindblad
generator. Although this setting should not be confused with a fundamental
model of wave-function collapse, it is useful to compare it with another
context in which conservation laws may fail, namely idealized
state-reduction processes~\cite{MinottiModanese2026Collapse}.

In a closed Schr\"odinger dynamics, local charge conservation follows from
the equations of motion and from the global gauge symmetry of the action.
If, however, the wave function is assumed to undergo an instantaneous
state reduction at time $t_0$,
\begin{equation}
    \psi
    =
    \psi_+\Theta(t-t_0)
    +
    \psi_-\Theta(t_0-t),
\end{equation}
then the usual Noether current need not be locally conserved during the
reduction event. In such an idealized description one obtains a
distributional source term of the form
\begin{equation}
    \partial_\mu J^\mu
    =
    q\left(
        |\psi_+|^2-|\psi_-|^2
    \right)\delta(t-t_0).
\end{equation}
This expresses the fact that the charge density changes discontinuously
between the pre-reduction and post-reduction states, without the
corresponding change being accounted for by the divergence of the
ordinary Schr\"odinger current.

The Lindblad collapse operators considered in the present work have a
different status. They are not assumed to represent a fundamental
instantaneous collapse of the wave function, but rather define an
effective Markovian evolution for the density matrix. Nevertheless, at
the level of expectation values, the dissipative part of the adjoint
Liouvillian plays a role analogous to a source term for the observable
under consideration. 

To make the analogy more explicit, consider a lattice or few-mode
system with local number operators $n_i$. If the Hamiltonian contains
coherent hopping terms between different sites or modes, the Hamiltonian
part of the evolution of $\langle n_i\rangle$ can often be written in
the form of a discrete continuity equation,
\begin{equation}
    \left.
    \frac{d}{dt}\langle n_i\rangle
    \right|_{H}
    =
    -\sum_{j\in \mathcal N(i)}
    \langle J^{(H)}_{i\to j}\rangle ,
\end{equation}
where $\mathcal N(i)$ denotes the set of sites or modes coupled to $i$,
and $J^{(H)}_{i\to j}$ is the Hamiltonian-generated current operator from $i$ to
$j$, with the convention $J^{(H)}_{i\to j}=-J^{(H)}_{j\to i}$.
The sum over $j$ is the lattice analogue of the divergence of a current:
it is the net current leaving site $i$ due to the Hamiltonian dynamics.

When dissipative Lindblad terms are added, the same local balance becomes
\begin{equation}
    \frac{d}{dt}\langle n_i\rangle
    +
    \sum_{j\in \mathcal N(i)}
    \langle J^{(H)}_{i\to j}\rangle
    =
    \left\langle
        \mathcal D^\dagger(n_i)
    \right\rangle .
    \label{eq:lattice_continuity_lindblad}
\end{equation}
The right-hand side is therefore an effective source or sink for the
local density $n_i$. If it vanishes, the local change of
$\langle n_i\rangle$ is entirely accounted for by Hamiltonian currents to
neighbouring sites. If it does not vanish, the dissipative dynamics
creates or removes occupation locally, in the same formal position as a
source term in a continuity equation.

Thus the condition
\begin{equation}
    \mathcal D^\dagger(n_i)=0
\end{equation}
is the Lindblad analogue of local conservation for the corresponding
density.

The comparison is especially useful when considering fluctuations. In
state-reduction models, event-level violations of local conservation may
cancel in the statistical average, while still leaving possible signatures
in fluctuations, correlations, or detector responses. The same distinction
appears naturally in the Lindblad framework. Even if
\begin{equation}
    \left\langle
        \mathcal D^\dagger(O)
    \right\rangle =0
\end{equation}
for a certain state or ensemble, the second moment may still evolve,
\begin{equation}
    \left\langle
        \mathcal D^\dagger(O^2)
    \right\rangle \neq 0.
\end{equation}
In that case the mean value of $O$ is conserved, but its variance is not.
This provides a simple open-system analogue of the distinction between
statistical conservation and event-level non-conservation.

\section{Conclusions}
\label{sec:conclusions}

We have analyzed the statistics of observables that are conserved by the
Hamiltonian dynamics of a closed quantum system but not necessarily by the
full Lindblad evolution. The basic point is that the Hamiltonian condition
$[H,O]=0$ is not sufficient in an open system: conservation of the
expectation value for all states requires the stronger condition
$\mathcal L^\dagger(O)=0$. When this condition fails, the dissipative part
of the adjoint Liouvillian acts as an effective source or sink for the
observable.

The same adjoint formalism gives an equally simple expression for the
second moment and for the variance. This is important because a dissipative
process may affect not only the mean value of a formerly conserved
quantity, but also the width of its distribution. Conversely, there can be
situations in which a mean value is conserved for symmetry or statistical
reasons while fluctuations still change, as explicitly shown by the
momentum-diffusion example in Sec.~\ref{sec:variance-only}. The pair of quantities
$\langle O\rangle$ and $\mathrm{Var}(O)$ therefore provides a minimal
statistical assessment of the breaking, preservation, or statistical
restoration of a conservation law.

The examples considered here illustrate this assessment in simple analytic
settings. In one-qubit amplitude damping, $\sigma_z$ is Hamiltonian
conserved but relaxes under the dissipator, while the second moment is
trivial because $\sigma_z^2=\mathbb I$. The two-qubit exchange model gives
a less degenerate example in which the total excitation number is conserved
by the coherent exchange Hamiltonian but is depleted by local decay, with
a nontrivial evolution of the second moment and variance. The
Jaynes--Cummings model provides a standard quantum-optical example: its
closed Hamiltonian conserves the total excitation number, but photon loss
and atomic decay destroy this conservation, whereas a pure-dephasing
reservoir coupled through the excitation number itself preserves the
statistics of that number.

Finally, we have emphasized that the Lindblad collapse operators used in
open-system master equations should not be identified with a fundamental
state-reduction process. Nevertheless, the adjoint dissipator provides a
natural effective source term for local observables, and this makes contact
with discussions of event-level non-conservation in idealized
wave-function-reduction models. This analogy is particularly useful at the
level of fluctuations: even when a source vanishes in the mean, second
moments and correlations may retain information about the underlying
stochastic non-conservation events.

\bibliographystyle{apsrev4-2}
\bibliography{references}

\end{document}